\documentclass[]{spie}  

\usepackage{color}
\usepackage{graphicx}
\usepackage{subfigure}
\usepackage{bm}
\usepackage{amsmath}
\usepackage{amssymb}
\usepackage{cite}
\usepackage{stfloats}
\usepackage{bm}
\usepackage{mathrsfs}

\usepackage{subfigure}
\usepackage{amsmath}
\usepackage{amssymb}
\usepackage{stfloats}

\title{Highly efficient quantitative phase microscopy using programmable annular LED illumination}

\author[]{Jiaji Li$^{1,2,3,4}$}
\author[]{Qian Chen$^{1,2}$}
\author[]{Jialin Zhang$^{1,2,3}$}
\author[]{Yan Zhang$^{1,2,3}$}
\author[]{Linpeng Lu$^{1,2,3}$}
\author[]{Chao Zuo$^{1,2,3,*}$}

\affil[]{$^1$School of Electronic and Optical Engineering, Nanjing University of Science and Technology, No. 200 Xiaolingwei Street, Nanjing, Jiangsu Province 210094, China\\
$^2$Jiangsu Key Laboratory of Spectral Imaging \& Intelligent Sense, Nanjing University of Science and Technology, Nanjing, Jiangsu Province 210094, China\\
$^3$Smart Computational Imaging (SCI) Laboratory, Nanjing University of Science and Technology, Nanjing, Jiangsu Province 210094, China\\
$^4$jiajili@njust.edu.cn\\}

\authorinfo{Further author information: (Send correspondence to C. Zuo)\\E-mail: surpasszuo@163.com}

\begin{document}
\maketitle

\begin{abstract}
In this work, we present a highly efficient quantitative phase imaging (QPI) approach using programmable annular LED illumination based on traditional bright-field microscope. As a new type of coded illumination, the LED array provides a flexible and compatible way to realize QPI. The proposed method modulates the transfer function of system by changing the LED illumination pattern, which achieves twice resolution of objective NA and gives noise-robust response of transfer function. The phase of a sample could be recovered from the intensity images with the inversion of transfer function. Moreover, the weak object transfer function (WOTF) of axisymmetric oblique source is derived, and the noise-free and noisy simulation results sufficiently validate the applicability  of discrete annular source. The quantitative phase measurements of micro polystyrene bead and visible blazed transmission grating demonstrate the accuracy of proposed method. Finally, the experimental investigations of unstained human cancer cells using different types objective are presented, and these results show the possibility of widespread adoption of QPI in the morphology study of cellular processes and biomedical community.
\end{abstract}

\section{Introduction}

Phase imaging plays a crucial role in the fields of optical, X-ray and electron microscopy \cite{optical1,optical2,x-ray,electron}. The phase of biological cells and tissues carries important information about the structure and intrinsic optical properties in microscopic imaging. Although this information cannot be directly recorded by the digital detector (CCD or CMOS), the Zernike phase contrast microscopy \cite{PCM} and differential interference contrast (DIC) microscopy \cite{DIC} could provide reliable phase contrast about the transparent cells and weakly absorbing objects via converting the phase into intensity. However, these techniques just only be used for the visualized and qualitative imaging instead of giving quantitative maps of phase change, which makes the quantitative data interpretation and phase reconstruction difficult.

Quantitative phase imaging (QPI) is a powerful tool for wide-ranging biomedical research and characterization of optical elements, which can realize the quantitative reconstruction of the sample information due to its label-free and unique capabilities to image the phase or the optical path thickness of cells, tissues, and optical fibers. As the conventional interferometric approach of QPI, off-axis digital holographic microscopy (DHM) \cite{DMH1,DMH2} measures the phase delay quantitatively introduced by the heterogeneous refractive index distribution within the specimen. Such method requires a coherent illumination source and a relatively complicated, vibration-sensitive optical system, even the speckle noise of laser degrades the spatial resolution of phase image. By contrast, other non-interferometric QPI approaches, which are based on common path geometries, utilizing white-light illumination \cite{wl1,wl2,wl3} have been developed to alleviate the problem of coherent noise and enhance the stability of mechanical vibrations, thus the spatial resolution and imaging quality of the phase measurement are greatly improved. Nevertheless, these quantitative phase measurements based on QPI techniques still rely on spatially coherent illumination, the maximum achievable resolution of phase imaging is only dependent on the  numerical aperture (NA) of objective and restricted by the coherent diffraction limit.

On the other hand, the deterministic phase retrieval can also be realized by the transport of intensity equation (TIE) \cite{TIE1,TIE2,TIE3} only using object field intensities at multiple axially displaced planes. The TIE linearizes the relationship between the phase and derivative of intensity along the axis of propagation \cite{TIE1}, then the direct phase can be uniquely determined by solving the TIE with intensity images and the longitudinal intensity derivative on the in-focus plane. QPI based on TIE has been increasingly investigated in micro-optics inspection and dynamic phase imaging of biological processes in recent years due to its unique advantages over interferometric techniques to achieve quantitative reconstruction result without the need for complicated interferometric optical configurations, reference beam, laser illumination sources and phase unwrapping \cite{TIE_Appl1,TIE_Appl2,TIE_Appl3}. It has been demonstrated that the non-interferometric phase retrieval methods based on TIE can be well adopted to partially coherent illumination \cite{PC_TIE0,PC_TIE1,PC_TIE2,PC_TIE3} in spite of the fact that the original derivation of TIE is established on the paraxial approximation and coherent illumination. Due to the non-linear relationship among the intensity image of object, illumination source, and optical system under partially coherent field, the imaging process and mathematical modeling become more complicated than coherent situation \cite{Partial_Con1,Partial_Con2}. Nevertheless, the phase retrieval of TIE could be reformulated informatively using of the concept of weak object transfer function (WOTF) under weak defocus assumptions and ignoring the bilinear terms originating from the self-interference of scattered light \cite{WOTF1,WOTF2,WOTF3,WOTF4}. The WOTF describes the frequency domain response of phase and absorption for a certain optical imaging system, which is also been called the contrast transfer function (CTF) in the field of propagation-based X-ray phase imaging \cite{CTF1,CTF2}.

Although the reconstructed phase from TIE is not well-defined over the partially coherent field, the definition of ``phase'' has been proven to be the weighted average superposition of phase under various coherent illumination using the theory of coherent mode decomposition \cite{Partial_decomp}, and can be converted to the well-defined optical path length of sample \cite{PC_TIE2}. The physical meaning of phase for partially coherent field is related to the transverse Poynting vector \cite{PC_TIE0} or Wigner distribution moment \cite{Winger} as well. Under the coherent illumination, the transfer function would be truncated by the limit of objective NA, and the poor response of TIE at low spatial frequency amplifies  the noise and leads the cloud-like artifacts superimposed on the reconstructed phases \cite{TIE3,TIE_Appl1}. While in the case of partially coherent light, the maximum achievable resolution of phase imaging is extended to the sum of objective NA and illumination NA over coherent situation, where the ratio of illumination NA to objective NA is called coherence parameter $s = N{A_{ill}}/N{A_{obj}}$. As the value of parameter $s$ increases ($N{A_{ill}} \le N{A_{obj}}$ actually), the phase contrast of defocused intensity image is vanished dramatically due to the attenuated response of transfer function. While the illumination NA approaches objective NA, the spatial cutoff frequency is increased to the two times of objective NA as predicted by the WOTF \cite{WOTF1,OTF2}, but meanwhile the low contrast intensity images would lead to the disadvantage that the signal-to-noise ratio (SNR) is too bad to recovery the phase from the defocused intensity images. The imaginary part of WOTF of large defocus distance rises faster than small defocus distance at low spatial frequency near zero frequency, so most of phase retrieval methods with TIE based on multiple defocus-plane select the low frequencies of large defocus as the optimal one \cite{WOTF4,PC_TIE3}. But, the transfer function of phase under large defocus distances contains too much zero-crossings due to the  rhythmical fluctuation of sine function, and these points make it almost impossible to recovery the high frequencies information of phase.

In this paper, we present a highly efficient QPI approach which combines the annular aperture and programmable LED illumination by replacing traditional halogen illumination source with a LED array within a conventional transmission microscope. The annular illumination pattern matched with objective pupil is displayed on the LED array and each isolated LED is treated as the coherent source. The WOTF of axisymmetric oblique source in arbitrary position on source pupil plane is derived and the principle of discrete annular LED illumination pattern is validated. Not only the spatial resolution of final reconstructed phase could be extended to 2 NA of objective, but also the phase contrast of defocused intensity image is strong because the response of phase transfer function (PTF) with annular source tends to be roughly constant across a wide range of frequencies, which is an ideal form for noise-robust, high-resolution, and well-posed phase reconstruction.

Even though this TIE-based QPI approach utilizing annular illumination has been reported by our group in the earlier paper \cite{AI_TIE} and the LED array has also been employed for Fourier ptychography \cite{FP1,FP2} and other QPI modalities \cite{QP_LED1,QP_LED2}, the novelty of this work is to deduce the WOTF for axisymmetric oblique source, and develop this discrete source to the superposition of arbitrary illumination pattern, such as circular illumination, annular illumination, or any other axisymmetric illumination. Furthermore, the combination of annular illumination and programmable LED array makes the modulation of illumination more flexible and compatible without the need for anodized and dyed circular glass plate or customized 3D printed annuli \cite{AI_TIE}. These advantages make it a competitive and powerful alternative to traditional bright-field illumination approaches for wide variety of biomedical investigations, micro-optics inspection and biophotonics. The noise-free and noisy simulation results sufficiently validate the applicability of discrete annular source, and the quantitative phase measurements of a micro polystyrene bead and visible blazed transmission grating demonstrate the accuracy of this method. The experimental investigation of unstained human cancer cells using different types objective are presented, and this results show the possibility of widespread adoption of QPI in the morphology study of cellular processes and biomedical community.

\section{Principle}
\subsection{WOTF for axisymmetric oblique source}
In the standard 6$f$ optical configuration, illustrated in Figure 1 of \cite{WOTF1}, an object is illuminated by the k\"ohler illumination source and imaged via an objective lens. The image formation of this telecentric microscopic imaging system could be described by the Fourier transform and a linear filtering operation in the pupil plane \cite{Partial_Con1}. For the incoherent case, the intensity image can be given by the convolution equation $I\left( \mathbf{r} \right)={{\left| h\left( \mathbf{r} \right) \right|}^{2}}\otimes {{\left| t\left( \mathbf{r} \right) \right|}^{2}}\text{=}{{\left| h\left( \mathbf{r} \right) \right|}^{2}}\otimes {{I}_{u}}\left( \mathbf{r} \right)$, where $h$ denotes the amplitude point spread function (PSF) of the imaging system, $t$ is the complex amplitude, and ${I_u}$ represents the intensities of coherent partial images arising from all light source points. On a different note, in the coherent case it obeys $I\left( \mathbf{r} \right) = {\left| {h\left( \mathbf{r} \right) \otimes t\left( \mathbf{r} \right)} \right|^2}$. Thus, the incoherent system is linear in intensity, whereas the coherent system is highly nonlinear in that quantity \cite{Partial_Con1}. More information about how to obtain the intensity under partially coherent illumination can be found in the Appendix A.

Due to the fact that above image formation is not linear in either amplitude or intensity, the mathematical derivation of phase recovery becomes more complicated for partially coherent system \cite{Partial_Con1,Partial_Con2}. To simplify this theoretical modeling, one way is to assume that the observed sample is a weak phase object, and the first-order Taylor expansion of complex amplitude can be described as:
\begin{equation}\label{1}
t\left( {\bf{r}} \right) \equiv a\left( {\bf{r}} \right)\exp \left[ {i\phi \left( {\bf{r}} \right)} \right] \approx a\left( {\bf{r}} \right){\left[ {1 + i\phi \left( {\bf{r}} \right)} \right]^{a\left( {\bf{r}} \right) = {a_0} + \Delta a\left( {\bf{r}} \right)}} \approx {a_0} + \Delta a\left( {\bf{r}} \right) + i{a_0}\phi \left( {\bf{r}} \right)
\end{equation}
where $a\left( {\bf{r}} \right)$ is the amplitude with a mean value of ${a_0}$, and $\phi \left( {\bf{r}} \right)$ is the phase distribution. Implementing the Fourier transform to $t$ and multiplying it with its conjugate form, the interference terms of the object function (bilinear terms) can be neglected owing to the scattered light is weak compared with the un-scattered light for a weak phase object. The formula of complex conjugate multiplication can be approximated as:
\begin{equation}\label{2}
\begin{aligned}
T\left( {{{\bf{u}}_1}} \right){T^*}\left( {{{\bf{u}}_2}} \right) =  & a_0^2\delta \left( {{{\bf{u}}_1}} \right)\delta \left( {{{\bf{u}}_2}} \right) + {a_0}\delta \left( {{{\bf{u}}_2}} \right)\left[ {\Delta \widetilde a\left( {{{\bf{u}}_1}} \right) + i{a_0}\widetilde \phi \left( {{{\bf{u}}_{{1}}}} \right)} \right]   \\
&+ {a_0}\delta \left( {{{\bf{u}}_1}} \right)\left[ {\Delta {{\widetilde a}^*}\left( {{{\bf{u}}_2}} \right) - i{a_0}{{\widetilde \phi }^*}\left( {{{\bf{u}}_2}} \right)} \right].
\end{aligned}
\end{equation}

The approximation used in Eq. (\ref{2}) corresponds to the first order Born's approximation, and this approximation is commonly used in optical diffraction tomography \cite{ODT0,ODT1}. While the two cross-related points are coinciding with each other in the frequency domain, the intensity image under the partially coherent field for a weak object can be rewritten as the following equation by substitute Eq. (\ref{2}) into Eq. (\ref{27}) in the Appendix A:
\begin{equation}\label{3}
I\left( {\bf{r}} \right) = a_0^2TCC\left( {0;0} \right) + 2{a_0}{\mathop{\rm Re}\nolimits} \left\{ {\int {TCC\left( {{\bf{u}};0} \right)\left[ {\Delta \widetilde a\left( {\bf{u}} \right) + i{a_0}\widetilde \phi \left( {\bf{u}} \right)} \right]\exp \left( {i2\pi {\bf{ru}}} \right)d{\bf{u}}} } \right\}
\end{equation}
where the $TCC^{\rm{*}}\left( {0;{\bf{u}}} \right)$ is equal to $TCC\left( {{\bf{u}};0} \right)$ due to the conjugate symmetry of transmission cross coefficient (TCC). The intensity contribution of various system components (eg. source and object) is separated and decoupled in Eq. (\ref{3}), and the $TCC\left( {{\bf{u}};0} \right)$ could be expressed as WOTF:
\begin{equation}\label{4}
WOTF\left( \mathbf{u} \right)\equiv TCC\left( \mathbf{u};0 \right)=\iint{S\left( {{\mathbf{u}}^{'}} \right)}P^*\left( {{\mathbf{u}}^{'}} \right)P\left( {{\mathbf{u}}^{'}}+\mathbf{u} \right)d{{\mathbf{u}}^{'}}
\end{equation}
where ${\mathbf{u}}^{'}$ represents the variable in Fourier polar coordinate. The WOTF is real and even as long as the distribution of source $S\left( {\bf{u}} \right)$ or objective pupil $P\left( {\bf{u}} \right)$ is axisymmetric, thus the intensity image on the in-focus plane gives no phase contrast but absorption contrast. Some other asymmetric illumination methods could produce the phase contrast in the in-focus intensity image by break the symmetry of $S\left( {\bf{u}} \right)$ or $P\left( {\bf{u}} \right)$, and the prominent examples are differential phase contrast microscopy\cite{Axisys,QP_LED1} and partitioned or programmable aperture microscopy \cite{Program_micro1,Program_micro2}. The defocusing of optical system along $z$ axial, which is another more convenient way to produce phase contrast and an imaginary part, would be introduced into the pupil function:
\begin{equation}\label{5}
P\left( {\bf{u}} \right) = \left| {P\left( {\bf{u}} \right)} \right|{e^{ikz\sqrt {1 - {\lambda ^2}{{\left| {\bf{u}} \right|}^2}} }}, \left| {\bf{u}} \right|\lambda  \le 1
\end{equation}
where $z$ is the defocus distance along the optical axis. Substituting the complex pupil function into Eq. (\ref{4}) yields a complex WOTF:
\begin{equation}\label{6}
WOTF\left( {\bf{u}} \right) = \iint{ S\left( {{{\bf{u}}^{'}}} \right)\left| {{P^*}\left( {{{\bf{u}}^{'}}} \right)} \right|\left| {P\left( {{{\bf{u}}^{'}} + {\bf{u}}} \right)} \right|\exp \left[ {ikz\left( { - \sqrt {1 - {\lambda ^2}{{\left| {{{\bf{u}}^{'}}} \right|}^2}} {\rm{ + }}\sqrt {1 - {\lambda ^2}{{\left| {{\bf{u}}{\rm{ + }}{{\bf{u}}^{'}}} \right|}^2}} } \right)} \right]d{{\bf{u}}^{'}}}
\end{equation}
The transfer functions of amplitude and phase component are corresponding to the real and imagery part of WOTF, respectively:
\begin{equation}\label{7}
\begin{aligned}
  & {{H}_{A}}\left( \mathbf{u} \right)=2{{a}_{0}}\operatorname{Re}\left[ WOTF\left( \mathbf{u} \right) \right] \\
 & {{H}_{P}}\left( \mathbf{u} \right)=2{{a}_{0}}\operatorname{Im}\left[ WOTF\left( \mathbf{u} \right) \right].
\end{aligned}
\end{equation}

\begin{figure}[!b]
    \centering
    \includegraphics[width=11.5cm]{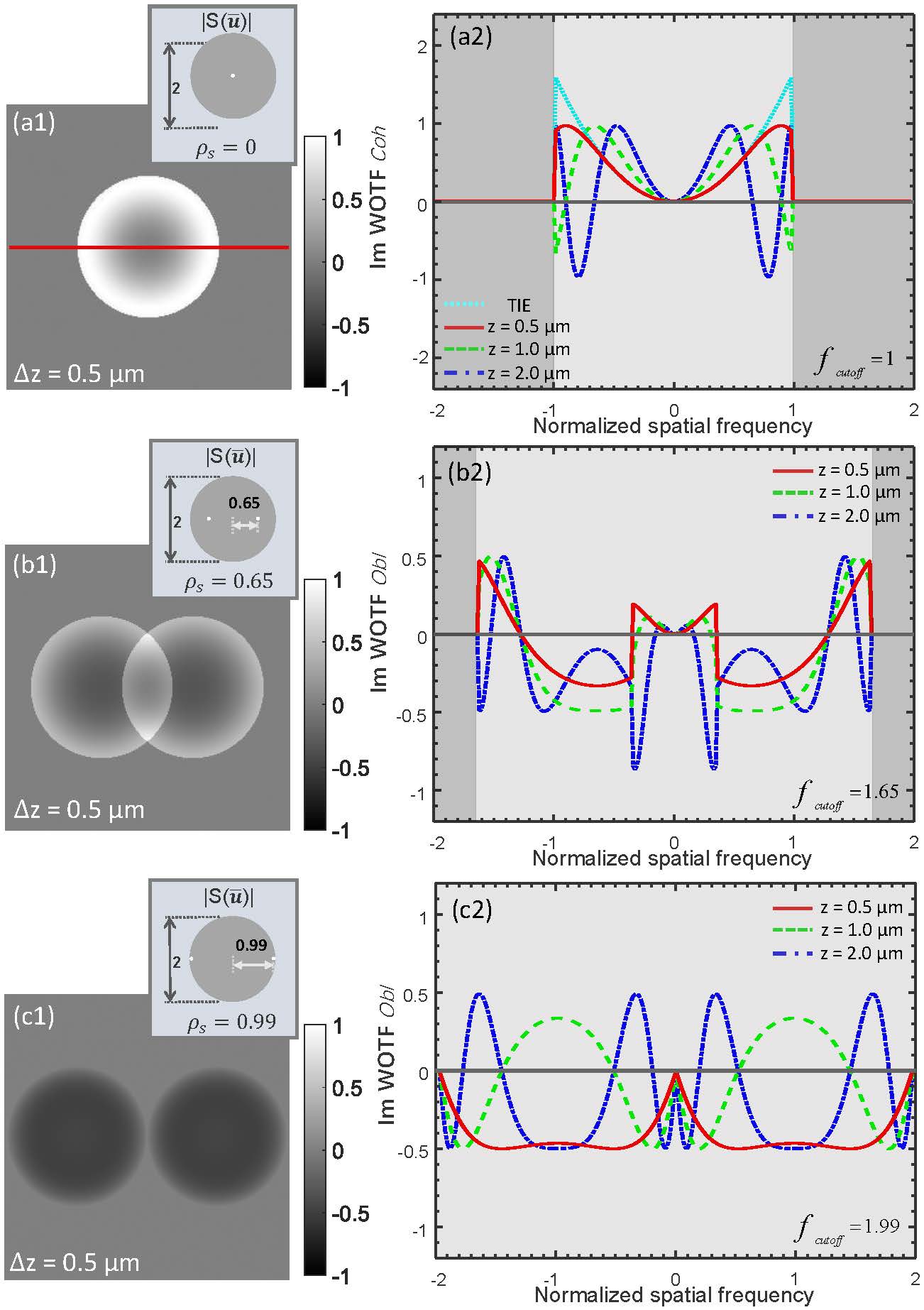}
    \caption{2D images of PTF for different types axisymmetric source under weak defocusing conditions and the line profiles of TIE and PTF for various defocus
distances.}
    \label{}
\end{figure}

Considering that the upright incident coherent source is a special case of oblique illumination, the derivation of WOTF for oblique source will be processed under the same framework for these two different types illumination. There is a pair of symmetrically ideal light spots on the source pupil plane, and the distance from these two points to the center point is ${\bm{\rho}_s}$ (normalized spatial frequency). The intensity distribution of this source pupil could be expressed as:
\begin{equation}\label{8}
S\left( \mathbf{u} \right)=\delta \left( \mathbf{u}-{{\bm{\rho }}_{s}} \right)\text{+}\delta \left( \mathbf{u}+{{\bm{\rho }}_{s}} \right)
\end{equation}
Substituting this source pupil function into Eq. (\ref{6}) results in a complex (but even) WOTF for oblique situation
\begin{equation}\label{9}
\begin{aligned}
WOT{{F}_{obl}}\left( \mathbf{u} \right)\text{=} & \left| P\left( \mathbf{u}-{{\bm{\rho }}_{{s}}} \right) \right|{{e}^{ikz\left( -\sqrt{1-{{\lambda }^{2}}{{\left| {{\bm{\rho }}_{{s}}} \right|}^{2}}}\text{+}\sqrt{1-{{\lambda }^{2}}{{\left| \mathbf{u}-{{\bm{\rho }}_{{s}}} \right|}^{2}}} \right)}} \\
& + \left| P\left( \mathbf{u}+{{\bm{\rho }}_{{s}}} \right) \right|{{e}^{ikz\left( -\sqrt{1-{{\lambda }^{2}}{{\left| {{\bm{\rho }}_{{s}}} \right|}^{2}}}\text{+}\sqrt{1-{{\lambda }^{2}}{{\left| \mathbf{u}+{{\bm{\rho }}_{{s}}} \right|}^{2}}} \right)}}
\end{aligned}
\end{equation}
where $\left| {P\left( {{\bf{u}} - {{\bm{\rho }}_{s}}} \right)} \right|$ and $\left| {P\left( {{\bf{u}} + {{\bm{\rho }}_{s}}} \right)} \right|$ are the pair of aperture functions shifted by the oblique coherent source in Fourier space. The aperture function for a circular objective pupil with normalized spatial radius ${{\bm{\rho }}_p}$ is given by

\begin{equation}\label{10}
\left| P\left( \mathbf{u} \right) \right|=
\left\{
\begin{aligned}
& 1,\quad \text{if }\mathbf{u}\le {{\bm{\rho }}_{p}} \\
& 0, \quad \text{if }\mathbf{u}>{{\bm{\rho }}_{p}}.
\end{aligned}
\right.
\end{equation}

In the coherent case (${{\bm{\rho }}_{{s}}}{\rm{ = }}0$), the WOTF can be greatly simplified as:
\begin{equation}\label{11}
WOT{{F}_{coh}}\left( \mathbf{u} \right)\text{=}\left| P\left( \mathbf{u} \right) \right|{{e}^{ikz\left( -1\text{+}\sqrt{1-{{\lambda }^{2}}{{\left| \mathbf{u} \right|}^{2}}} \right)}}.
\end{equation}
The two aperture functions are overlapped each other in this situation, so the values of final coherent WOTF is only half. The absorption contrast and phase contrast are given by the real and imaginary parts of $WOT{F_{coh}}$ using Euler's formula as shown in Eq. (\ref{7}). By further invoking the paraxial approximation and replacing $\sqrt {1 - {\lambda ^2}{{\bf{u}}^2}} $ with $1 - {{{\lambda ^2}{{\bf{u}}^2}} \mathord{\left/{\vphantom {{{\lambda ^2}{{\bf{u}}^2}} 2}} \right.
 \kern-\nulldelimiterspace} 2}$, the imaginary part of the $WOT{F_{coh}}$ could be written as a sine term $\sin \left( {\pi \lambda z{{\left| {\bf{u}} \right|}^2}} \right)$. Under the condition of weak defocusing, this transfer function can be further approximated by a parabolic function
 \begin{equation}\label{12}
 {{H}_{p}}{{\left( \mathbf{u} \right)}_{TIE}}\text{=} \left| P\left( \mathbf{u} \right) \right| \sin \left( {\pi \lambda z{{\left| {\bf{u}} \right|}^2}} \right) \approx \left| P\left( \mathbf{u} \right) \right| \pi \lambda z{\left| {\bf{u}} \right|^2}
 \end{equation}\label{12}
This Laplacian operator is corresponding to the PTF of TIE in Fourier domain, and the two dimensional (2D) image of WOTF for coherent source under weak defocusing condition is shown in Fig. 1(a1). The line profiles of TIE and PTF for various defocus distances are illustrated in Fig. 1(a2) as well. It is obvious that the transfer function profile of TIE is consistent with the PTF for weak defocus distance (0.5 $\mu$m) at low frequency, so the coherent transfer function is getting closer to the  TIE as long as the defocus distance is getting smaller. In other words, the TIE is a special case of coherent transfer function under weak defocusing.

On the other hand, these two coherent points do not coincide with each other in the center of source plane, as shown in Fig. 1(b1) and (c1). The imaginary part of Eq. (\ref{9}) is limited by their own pupil functions, thus the PTF for oblique point source could be written as:
\begin{equation}\label{13}
\begin{aligned}
{{H}_{p}}{{\left( \mathbf{u} \right)}_{obl}}\text{=}& \frac{1}{2} \left| P\left( \mathbf{u}-{{\bm{\rho }}_{{s}}} \right) \right|\sin \left[ kz\left( \sqrt{1-{{\lambda }^{2}}{{\left| \mathbf{u}-{{\bm{\rho }}_{{s}}} \right|}^{2}}}-\sqrt{1-{{\lambda }^{2}}{{\left| {{\bm{\rho }}_{{s}}} \right|}^{2}}} \right) \right] \\
& + \frac{1}{2}\left| P\left( \mathbf{u}+{{\bm{\rho }}_{{s}}} \right) \right|\sin \left[ kz\left( \sqrt{1-{{\lambda }^{2}}{{\left| \mathbf{u}+{{\bm{\rho }}_{{s}}} \right|}^{2}}}-\sqrt{1-{{\lambda }^{2}}{{\left| {{\bm{\rho }}_{{s}}} \right|}^{2}}} \right) \right]
\end{aligned}
\end{equation}
Figure 1(b2) and (c2) show the curves of PTF for different ${{\bm{\rho }}_s}$ and defocus distances additionally. The cutoff frequency of transfer function is determined by the shifted aperture functions, and the achievable imaging resolution, which is equal to ${{\bm{\rho }}_{{p}}}{\rm{ + }}{{\bm{\rho }}_{{s}}}$, becomes bigger and bigger with the increment of ${{\bm{\rho}}_{s}}$  in the oblique direction. Nevertheless, the profile line of transfer function has two jump edges due to the overlap and superposition of two shifted objective pupil functions. The jump edge would induce zero-crossings and make the response of frequency bad around these points, thus these jump edges should be avoided as much as possible. While this pair of points source matches objective pupil (${{\bm{\rho }}_{{p}}}{\rm{ \approx }}{{\bm{\rho }}_{{s}}}$), not only the cutoff frequency of PTF could be extended to the twice resolution of coherent diffraction limit but also the frequency response of PTF is roughly constant in a specific direction under this axisymmetric oblique illumination.

\subsection{Validation of discrete annular LED illumination}

\begin{figure}[!b]
    \centering
    \includegraphics[width=12.5cm]{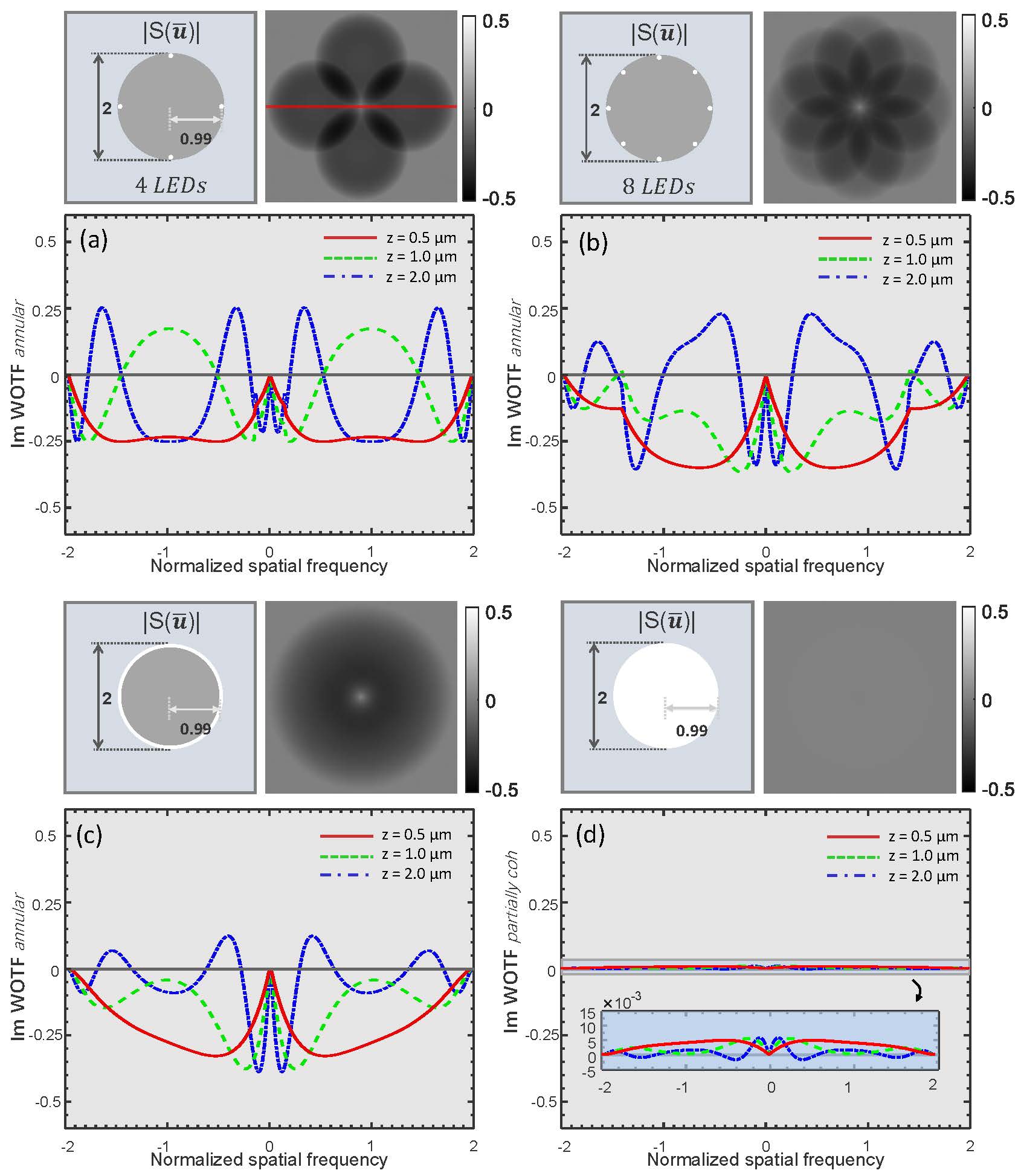}
    \caption{(a-c) 2D images of PTF and line profiles of three different types discrete annular illumination patterns for various defocus distances. (d) Traditional circular diaphragm aperture and corresponding PTF.}
    \label{}
\end{figure}

For any axisymmetric shape of partially coherent illumination, a certain illumination pattern could be discretized into a lot of coherent point source with the finite physical size including oblique and upright incident light points. The image formation of an optical microscopic system under partially coherent field could be simply understood as a convolution with a magnified replica of each discrete coherent source. Moreover, this process is coincident with the incoherent superposition of all intensities of the coherent partial images arising from all discrete light source points for the optical imaging with K\"ohler illumination \cite{Partial_Con1,Partial_decomp}. As we know, while the condenser aperture iris diaphragm becomes bigger, the maximum achievable imaging resolution of intensity image is getting bigger also and the depth of field (DOF) becomes shallower. However, the phase contrast (as well as absorption contrast) of the defocused image will become weak, and the attenuation of phase effect of captured intensity image will reduce the SNR of the phase reconstruction while the coherence parameter $s$ continues to grow \cite{WOTF1,AI_TIE}. So the parameter $s$ is recommended to be set between 0.7 and 0.8 for properly image resolution and contrast in most microscope instruction manual.

To overcome the tradeoff between image contrast and resolution, we present the highly efficient programmable annular illumination which is different from the traditional circular diaphragm aperture for QPI microscopy. The LED array is placed at the front focal plane of the condenser to illuminate the specimen, and each single LED could be controlled separately. A test image, which is used to simulate the discrete LED array, with 512 $\times$ 512 pixels with a pixel size of 0.176 $\mu$m $\times$ 0.176 $\mu$m and an objective with 0.75 NA  are employed for the validation of annular LED illumination. While a pair of oblique illumination points is located on the edge of source pupil, it could be known that the imaging resolution is twice objective NA in oblique direction as shown in Fig. 1(c). Thus, three different types of discrete annular patterns and one circular pattern are utilized for the comparison of WOTF under same system parameter. The expression of annular source could be written as the summation of delta function
\begin{equation}\label{14}
S({\bf{u}}) = \sum\limits_{i = 0}^N {\delta (\bf{u} - {{\bf{u}}_i})},\quad \left| {{{\bf{u}}_i}} \right| \approx \left| {{{\bm{\rho }}_p}} \right|
\end{equation}
where $N$ is the number of all discrete light points on the source plane.

Figure 2 shows the 2D images and line profiles of imaginary part of WOTF for various annular illumination patterns and defocus distances. There are four LEDs on the top-bottom and left-right of source plane in Fig. 2(a), so the double imaging resolution of objective NA could be obtained in the vertical and horizontal directions. While eight LEDs could cover the twice cutoff frequency of objective in four different directions, and the PTF image of eight LEDs seems to be the superposition of transfer function of several pairs axisymmetric oblique source. For the continuous situation of annular illumination, as shown in Fig.2(c), the final PTF provides isotropic imaging resolution in all directions. In addition to above three different types of annular shape, the PTF of circular illumination aperture is illustrated in Fig. 2(d) and the cutoff frequency is extended to 2 NA of objective as well. However, the value of transfer function of circular apertures is diminished dramatically compared to above three annular shapes. It is corresponding to the phenomenon that the larger aperture diaphragm provides higher imaging resolution but the phase contrast of defocused image is too weak to capture. The condenser aperture of circular illumination must be stopped down to produce appreciable contrast for phase information, but it is not necessary for the annular illumination. Here it is worth noting that the number of LED located on the edge of source pupil $N$ should be as much as possible for isotropic imaging resolution in all directions, but we chose the eight LEDs as the proposed illumination pattern considering the finite spacing between two adjacent LED elements.

From the plot of PTF for various aperture shapes and defocus distances, all four illumination patterns have twice frequency bandwidth of objective NA, but the response of circular illumination is too weak. The phase information can hardly be transferred into intensity via defocusing when illumination
NA is large, and the weak phase contrast of defocus intensity image would leads bad SNR. The zero-crossings number of PTF for large defocus distances is more than weak defocusing due to the rhythmical fluctuation of imaginary part of WOTF, and it is difficult to recovery the signal component from the noise around these point. Thereby, the proposed annular LED illumination pattern not only extends the imaging resolution to double NA in most directions but also provides the robust phase contrast response for defoused intensity image.

\subsection{QPI via TIE and WOTF inversion }

In the paraxial regime, the wave propagation is mathematically described by the Fresnel diffraction integral \cite{Partial_Con1}, while the relationship between the intensity and phase during wave propagation can be described by TIE \cite{TIE1}:
\begin{equation}\label{15}
-k\frac{\partial{I(\bm{r})}}{\partial{z}} = \nabla_\perp\bm\cdot[I(\bm{r})\nabla_\perp\phi(\bm{r})]
\end{equation}
where $k$ is the wave number ${2\pi }/{\lambda }$, $I(\bm{r})$ is the intensity image on the in-focus plane, $\nabla_\perp$ denotes the gradient operator over the transverse direction $\bm{r}$, $\bm\cdot$ denotes the dot product, and $\phi(\bm{r})$ represents the phase of object. The left hand of TIE is the spatial derivative of intensity on the in-focus plane along $z$ axis. The longitudinal intensity derivative $\partial{I}/\partial{z}$ can be estimated through difference formula ${\left( {{I_1} - {I_2}} \right)}$\slash${2\Delta z}$, where $I_1$ and $I_2$ are the two captured defocused intensity images, and $\Delta z$ is the defoucs distance of axially displaced image. By introducing the Teague's auxiliary function $\nabla_\perp\psi(\bm{r}) = I(\bm{r})\nabla_\perp\phi(\bm{r})$, the TIE is converted into the following two Poisson equations:
\begin{equation}\label{16}
-k\frac{\partial{I(\bm{r})}}{\partial{z}} = {\nabla_\perp}^2\psi
\end{equation}
and
\begin{equation}\label{17}
\nabla_\perp\bm\cdot(I^{-1}\nabla_\perp\psi) = {\nabla_\perp}^2\phi
\end{equation}
The solution for $\psi$ could be obtained by solving the first Poisson equation Eq. (\ref{16}), thus the phase gradient can be obtained as well. The second Poisson equation Eq. (\ref{17}) is used for phase integration, and the quantitative phase $\phi(\bm{r})$ can be uniquely determined by these two Poisson equations. For a special case of pure phase object (unstained cells and tissues generally), the intensity image on the in-focus plane could be treated as a constant because of the untainted cells is almost transparent, and the TIE can be simplified as only one Poisson equation:
\begin{equation}\label{18}
- k\frac{{\partial I\left( {\bf{r}} \right)}}{{\partial z}} = I\left( {\bf{r}} \right){\nabla ^2}\phi \left( {\bf{r}} \right)
\end{equation}
Then, the fast Fourier transform (FFT) solver \cite{TIE_Appl2,TIE_Appl3} is applied to Eq. (\ref{18}) and the forward form of TIE in the Fourier domain corresponds to a Laplacian filter
\begin{equation}\label{19}
\frac{{{{ \widetilde{I_1} }}\left( {\bf{u}} \right) - {{ \widetilde{I_2} }}\left( {\bf{u}} \right)}}{4{ \widetilde{I} \left( {\bf{u}} \right)}}  = \left( { \pi \lambda z{{\left| {\bf{u}} \right|}^2}} \right)\widetilde{\phi}(\bf{u})
\end{equation}
The inverse Laplacian operator $1\slash{\pi \lambda z{{\left| {\bf{u}} \right|}^2}}$ is analogous to an inversion of weak defocus CTF or PTF in the coherent limit.

\begin{figure}[!b]
    \centering
    \includegraphics[width=13.5cm]{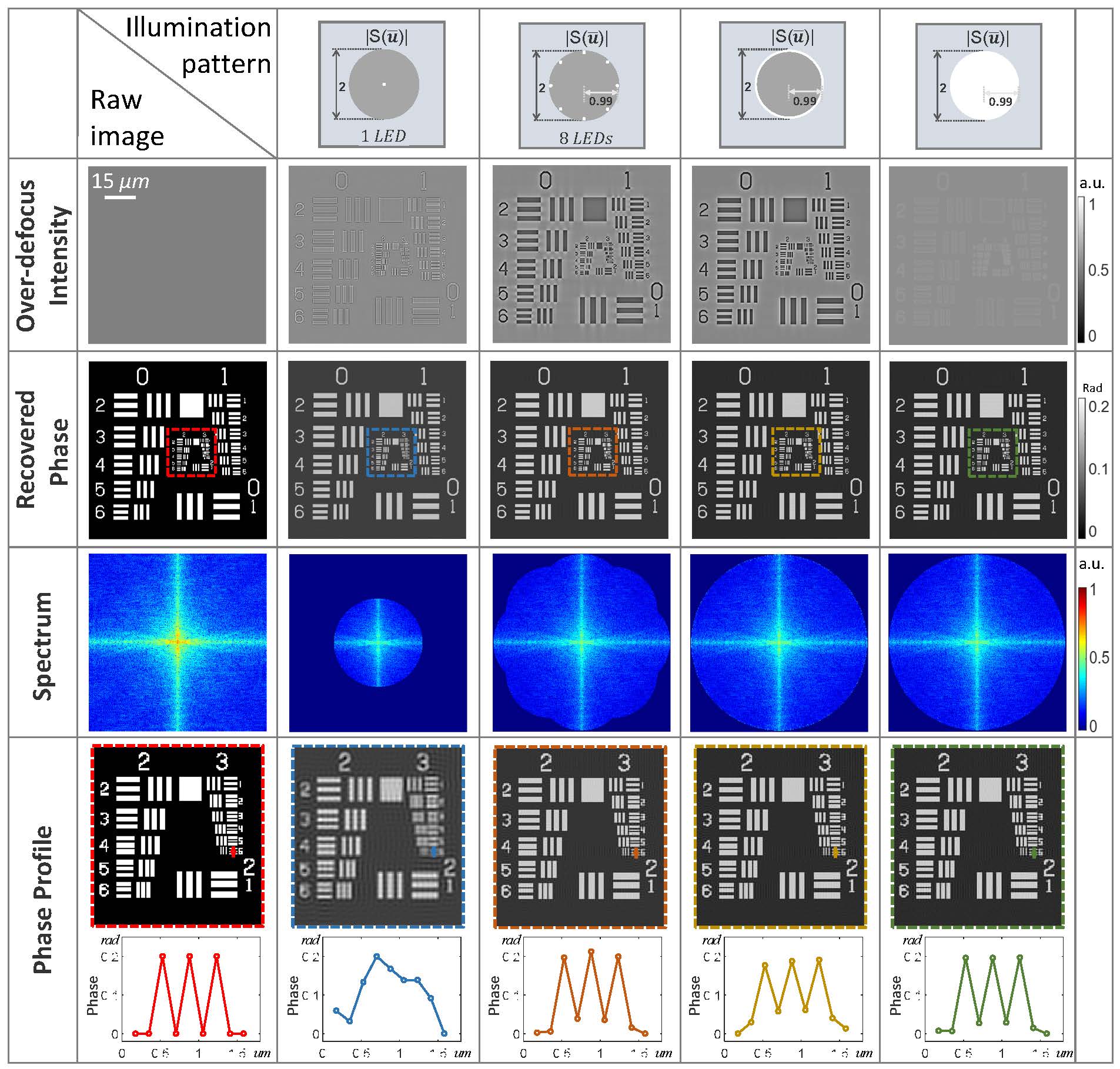}
    \caption{Various noise-free reconstruction results based on a simulated phase resolution target corresponding different illumination patterns. The parameters of optical system and pixel size of camera is set to satisfy the Nyquist sampling criterion, and the sampling frequency of camera equals twice imaging resolution of objective NA. Scale bar, 15 $\mu$m.}
    \label{}
\end{figure}

For partially coherent illumination, the traditional form of TIE is not suitable for the phase retrieval since this equation contains no parameters about imaging system. To take the effect of partial coherence and imaging system into account, the Laplacian operator $\pi \lambda z{\left| {\bf{u}} \right|^2}$ of TIE in the Fourier space should be replaced by the PTF of arbitrary axisymmetric source. The ATF ${H_A}\left( {\bf{u}} \right)$ and PTF ${H_{\rm{P}}}\left( {\bf{u}} \right)$ are determined by the real and imagery part of WOTF respectively, as shown in Eq. (\ref{7}). Thus, the ATF is an even function due to the nature of the cosine function, while the PTF is always an odd function for various different defocus distance. On the condition that the defoucs distances of two captured intensity images are same and defocus direction is opposite, the subtraction between two intensity images give no amplitude contrast but a pure twice phase contrast. Therefore, the in-focus image ${I(\bm{r})}$ is treated as the background intensity and the forward form of WOTF can be expressed as:
\begin{equation}\label{20}
\frac{{{{ \widetilde{I_1} }}\left( {\bf{u}} \right) - {{ \widetilde{I_2} }}\left( {\bf{u}} \right)}}{4{ \widetilde{I} \left( {\bf{u}} \right)}}  =  {\mathop{\rm Im}\nolimits} \left[ {WOTF\left( {\bf{u}} \right)} \right] \widetilde{\phi}(\bf{u})
\end{equation}
Equation (\ref{20}) makes the relationship between phase and PTF linear, then QPI can be realized by the inversion of WOTF in Fourier space
\begin{equation}\label{21}
\phi \left( {\bf{r}} \right) = {{\mathscr{F}}^{ - 1}} \left\{{ \frac{{{{ \widetilde{I_1} }}\left( {\bf{u}} \right) - {{ \widetilde{I_2} }}\left( {\bf{u}} \right)}}{4{ \widetilde{I} \left( {\bf{u}} \right)}}  {\frac{{{\mathop{\rm Im}\nolimits} \left[ {WOTF\left( {\bf{u}} \right)} \right]}}{{{{\left| {{\mathop{\rm  Im}\nolimits} \left[ {WOTF\left( {\bf{u}} \right)} \right]} \right|}^2} + \alpha }}}  } \right\}
\end{equation}
where ${{\mathscr{F}}^{ - 1}}$ denotes the inverse Fourier transform, and $\alpha$ is the Tikhonov-regularization parameter, which is usually used in the Wiener filter to set maximum amplification, avoiding the division by zero of WOTF.

First, we implement our method to the phase reconstruction of a simulated resolution target. The resolution test image is used as an example phase object defined on a square region and the grid width is 512 pixels with a pixel size of 0.176 $\mu$m. The wavelength of illumination is 530 nm, and the objective NA is 0.75. The captured defocused intensity images are noise-free and the defocus distance is 0.5 $\mu$m. The WOTF for various illumination patterns could be derived using Eq. (\ref{9}) and Eq. (\ref{11}), and the inversion of WOTF is applied to the Fourier transform of captured intensity stack. The detailed compare reconstruction results of resolution target under different illumination patterns are shown in Fig. 3. The NA of objective and the pixel size of camera are set to satisfy the Nyquist sampling criterion, and twice imaging resolution of objective NA is equal to the maximum sampling frequency of camera. The center region of simulated resolution target is enlarged and marked with the dashed rectangle. As predicted by the WOTF of corresponding illumination pattern, the recovered spectrum is determined by the cutoff frequency of WOTF. Also, the phase profile lines of resolution elements in the smallest Group in this simulated resolution test image are plotted in the last row of sub-figure, and it could be seen that the edge of the resolution elements of coherent illumination is distorted and blurry but the elements of other groups of three aperture patterns are distinguishable.

In order to characterize the noise sensitivity of proposed method, another simulated result is presented as well. The system parameters are same as above simulation, but each defocused intensity image is corrupted by Gaussian noise with standard deviation of 0.1 to simulate the noise effect. The shape of reconstructed Fourier spectrum is same as the non-zero region of PTF, and the final retrieved phase is evaluated by the root-mean-square error (RMSE). From this diagram, the cutoff frequency of coherent illumination is restricted to coherent diffraction limit, but the other three groups of source could extend the cut-off frequency to double imaging resolution of objective NA. Although the coherent situation could provides the maximum value of PTF (unit 1 approximatively), the slowly rising of PTF response at low frequency leads the over-amplification of noise, and the cloud-like artifacts is superimposed on the finally reconstructed phase. The values of WOTF of traditional circular aperture is too close to zero and results the over-amplification of noise at both low and high frequency. Therefore, the proposed annular illumination method provides not only the twice resolution of objective NA  but also the robust response of transfer function, and the accuracy and stable quantitatively retrieved phase of the test object is given at last.

\begin{figure}[!htp]
    \centering
    \includegraphics[width=13.5cm]{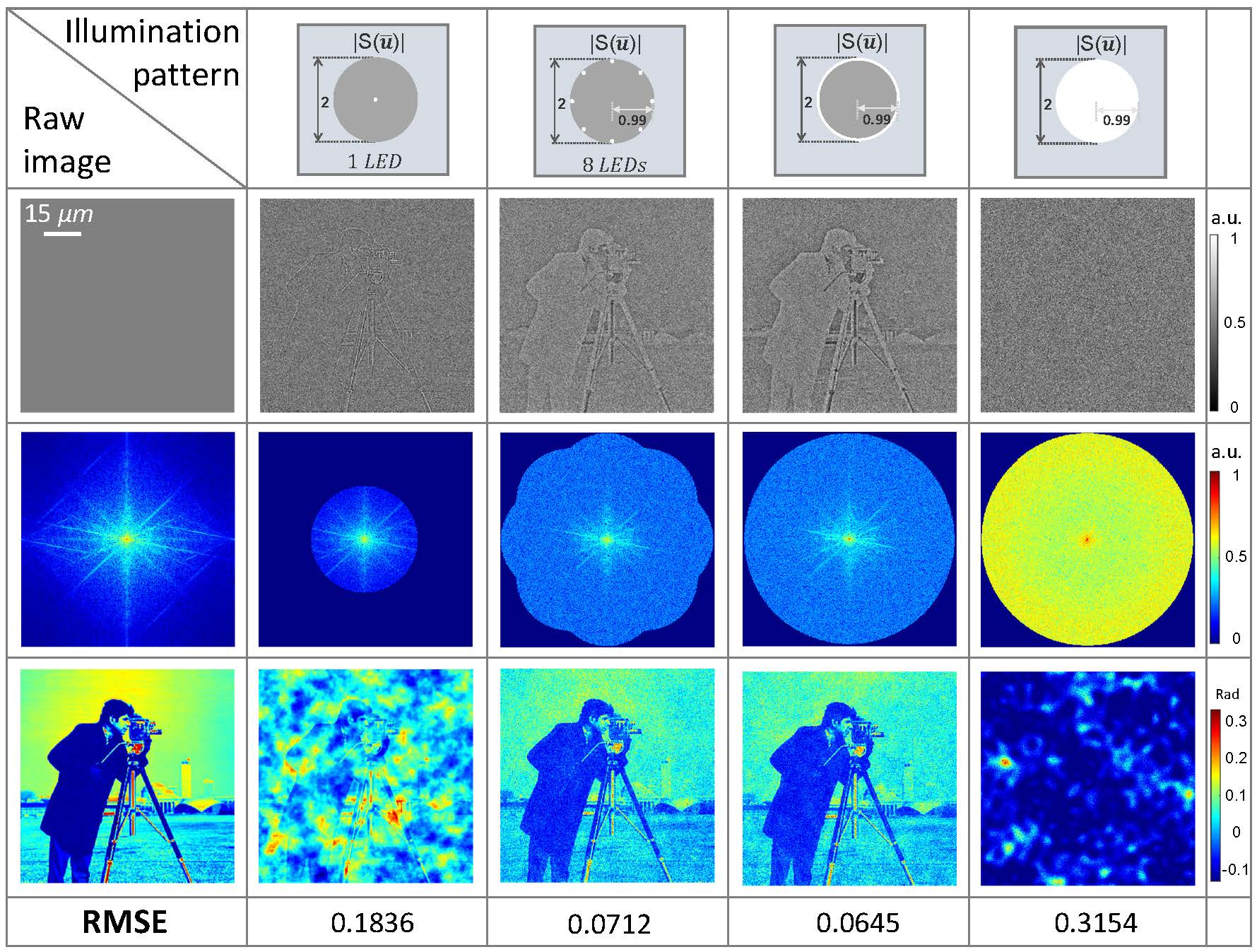}
    \caption{Phase reconstruction results under the Gaussian noise with standard deviation of 0.1. The response of transfer function rises slowly at low frequencies leading the over-amplification of noise, and there are cloud-like artifacts superimposed on the reconstructed phases for coherent illumination. While the values of WOTF of traditional circular aperture is too close to zero and leads the over-amplification of noise at both low and high frequency. Scale bar, 15 $\mu$m.}
    \label{}
\end{figure}

\section{Experimental setup}

\begin{figure}[!htp]
    \centering
    \includegraphics[width=13.5cm]{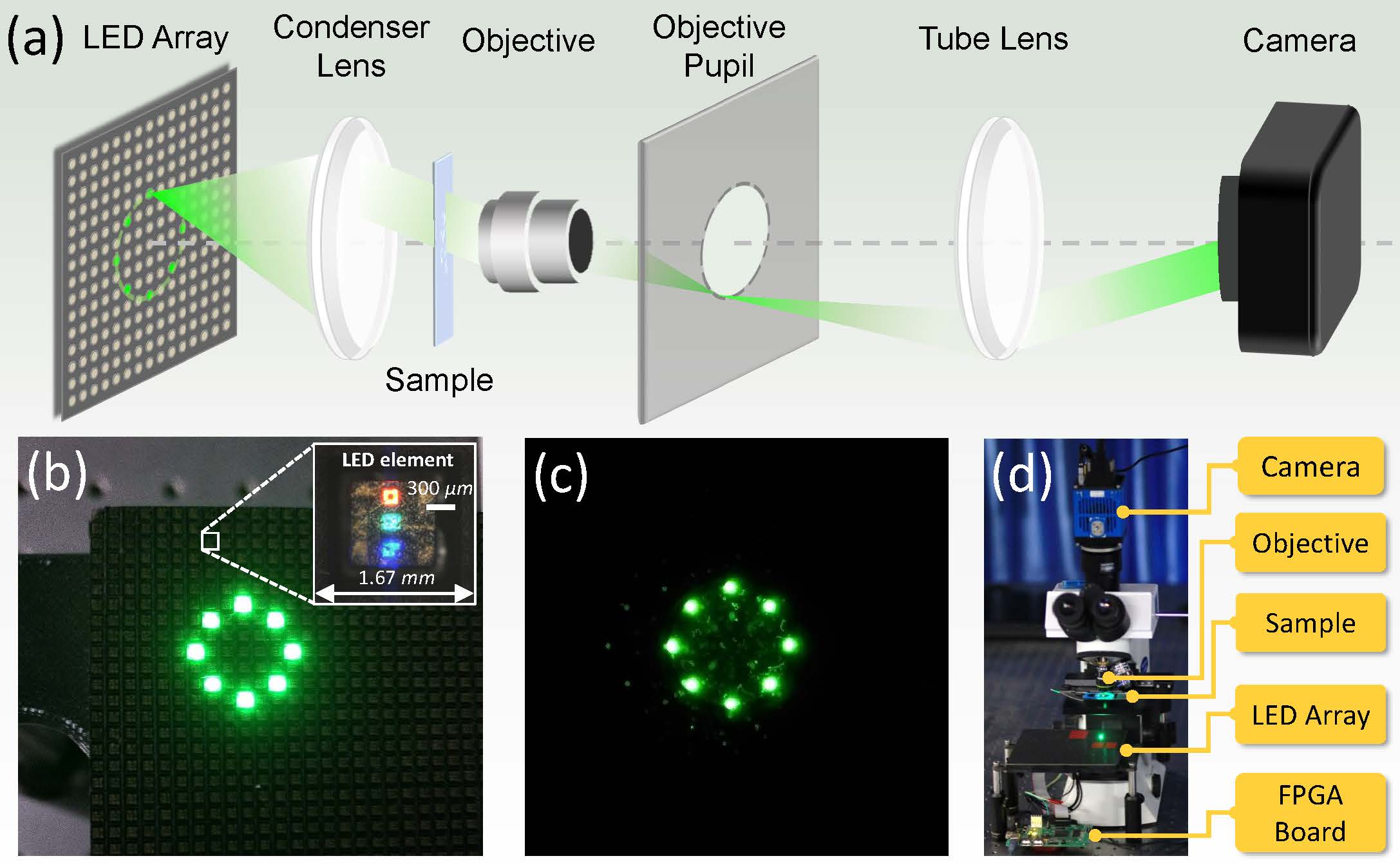}
    \caption{(a) Schematic diagram of highly efficient quantitative phase microscopy. (b-c) The annular pattern is displayed on the LED array and the size of this annulus is matched with objective pupil in the back focal plane. (d) Photograph of whole imaging system. The LED array is placed beneath the sample and the crucial parts of setup in this photo are marked with the yellow boxes. Scale bar represents 300 $\mu$m.}
    \label{}
\end{figure}

As depicted in Fig. 5(a), the highly efficient quantitative phase microscopy is composed of three major components: a programmable LED array, a microscopic imaging system, and a CMOS camera. The commercial surface-mounted LED array is placed at the front focal plane of the condenser as illumination source, and the light emitted from condenser lens for single LED can be nearly treated as a plane wave. Each LED can provide approximately spatially coherent quasi-monochromatic illuminations with narrow bandwidth (central wavelength $\lambda$ = 530 nm, $\sim$ 20 nm bandwidth). The distance between every adjacent LED elements is 1.67 mm, and only a fraction of whole array are used for programmable illumination. The array is driven dynamically using a LED controller board, which is custom-built by ourselves with a Field Programmable Gate Array (FPGA) unit, to provide the various illumination patterns.

In our work, the discrete annular LED illumination pattern matched with objective pupil is displayed on the array, as shown in Fig. 5(b). Figure 5(c) is taken in the objective back focal plane by inserting a Bertrand lens into one of the eyepiece observation tubes or removing the eyepiece tubes. The microscope is equipped with a scientific CMOS (sCMOS) camera (PCO.edge 5.5, 6.5 $\mu$m pixel pitch) and an universal plan objective (Olympus, UPlan 20 $\times$, NA = 0.4). Another universal plan super-apochromat objective (Olympus, UPlan SAPO 20$\times$, NA $=$ 0.75) and a higher sampling rate detector (2.2 $\mu$m pixel pitch) are also utilized for higher resolution imaging result. The photograph of whole imaging system is illustrated in Fig. 5(d) and the crucial parts of setup in this photo are marked with the yellow boxes.

\section{Results}
\subsection{Quantitative characterization of control samples}

\begin{figure}[!t]
    \centering
    \includegraphics[width=13cm]{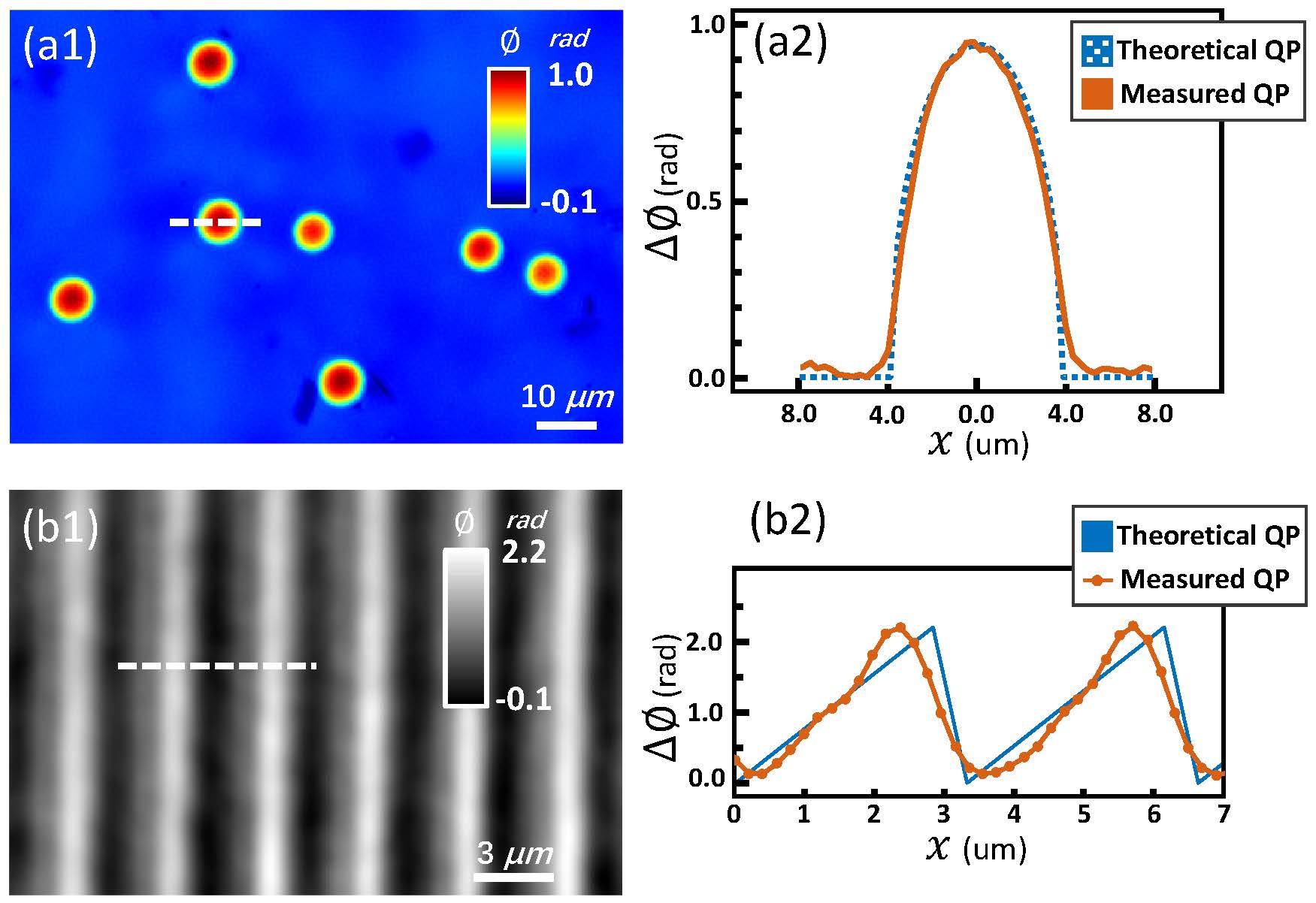}
    \caption{(a1-b1) Reconstructed phase distributions of the micro polystyrene bead with 8 $\mu$m diameter and blazed transmission grating with 3.33 $\mu$m period. (a2-b2) Measured quantitative phase line profiles for a single bead and a few periods grating. Theoretical (assuming 90$^\text{o}$ groove angles) line profiles are also plotted for reference. Scale bar denotes 10 $\mu$m and 3 $\mu$m, respectively.}
    \label{}
\end{figure}

In order to validate the accuracy of proposed QPI approach based on annular LED illumination, the micro polystyrene bead (Polysciences, $n$=1.59) with 8 $\mu$m diameter immersed in oil (Cargille, $n$=1.58) is measured using 0.4 NA objective and sCMOS camera. The sample is slightly defocused, and three intensity images are recorded at $\pm$ 1 $\mu$m plane and in-focus plane. By invoking the inversion of WOTF, the reconstructed quantitative phase image of bead is shown in Fig. 6(a1), which is a sub-region of whole field of view (FOV). The horizontal line profile through the center of a single bead is illustrated as the solid brown line in Fig. 6(a2), and the blue dash line represents the theoretical quantitative phase of the micro polystyrene bead. Of interest in these results is excellent agreement between the magnitude and shape of the compared bead profile. There is still some slight high frequency noise in the retrieved phase image due to the fact that the tiny value of WOTF amplifies the noise near the cutoff frequency, but these artifacts do not affect the accuracy and feasibility of our proposed method.

Further more, a visible blazed transmission grating ($Thorlabs\;GT13-03$, grating period $\Lambda$ = 3.33 $\mu$m, blaze angle ${\theta _B}$ = 17.5$^\text{o}$) is employed in the quantitative experiment using the same method and procedures. The grating is made by Schott B270 glass ($n_{glass}$ = 1.5251), and mounted face up on a glass slide with refractive index matching water ($n_{water}$ = 1.33) and a thin $no$. 0 coverslip. Considering that the large pixel size of sCMOS camera and high density of grating, a higher NA objective (NA = 0.75) and sampling rate detector (2.2 $\mu$m pixel size) are utilized for the imaging of this grating. The measured phase image is represented in Fig. 6(b1) for a 23.7 $\mu$m $\times$ 15.6 $\mu$m rectangular patch. Plotted for reference are the theoretical profiles in blue solid line, assuming 90$^\text{o}$ groove angles, and also plotted in Fig. 6(b1) is a few periods of the associated brown dot-solid line profiles with no interpolation. These two curves are well consistent with each other excepting the jump edges of phase owing to the rapid oscillations of grating. Thus, the two group quantitative characterizations of control samples further indicate success and accuracy of our method.

\subsection{Experimental results of biological specimens}

\begin{figure}[!htp]
    \centering
    \includegraphics[width=13cm]{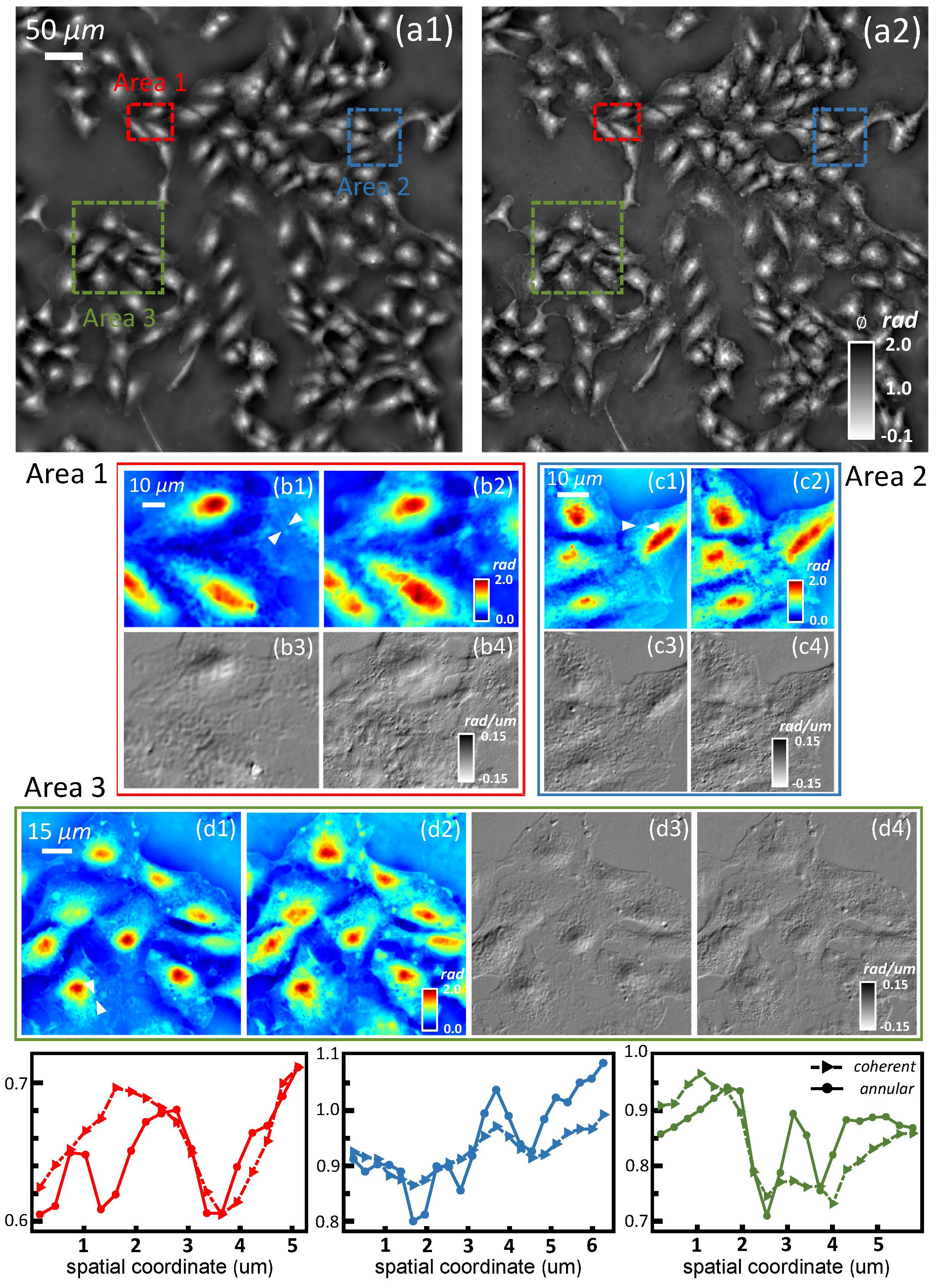}
    \caption{(a) Quantitative reconstruction results of LC-06 with 0.4 NA objective and 6.5 $\mu$m pixel pitch camera for coherent and discrete annular illumination. (b-c) Three enlarged sub-regions of quantitative maps and simplified DIC images are illustrated as well. The white arrows shows line profiles taken at different positions in the cells. Scale bar equals 50$\mu$m, 10$\mu$m and 15$\mu$m, respectively.}
    \label{}
\end{figure}

As demonstrated by the previous simulation results in subsection 2.3, the developed annular LED illumination could provides twice imaging resolution of objective NA and noise robust response of WOTF. We also test the present reconstruction method in its intended biomedical application experimentally, and the unstained lung cancer cell (LC-06) is used for highly efficient QPI firstly with 0.4 NA objective and 6.5 $\mu$m pixel pitch camera. Figure 7(a1) and (a2) are the quantitative phase images of LC-06 defined on a square FOV for point source and annular source respectively. Three representative sub-areas of whole quantitative map are selected and enlarged for more detailed descriptions. The phase images of three enlarged sub-regions are shown in jet map, and the corresponding simplified DIC images are illustrated in Fig. 7(b) and (c).

From these quantitative phase and phase gradient images, it is obvious that the phase imaging resolution of annular illumination source is higher than the coherent one, and some tiny grains in cytoplasm could be observed clearer and more vivid. In addition, the white arrows show line profiles taken from two different positions in the cells, and the comparative phase profiles are presented in different colors of the lines in Fig. 7. The plot lines indicate that the significant improvement of high frequency features using annular aperture as compared to the coherent illumination. Thus, the allowed highest spatial frequency of QPI base on annular LED illumination is 0.8 NA (0.66$\mu$m) effectively in the phase reconstruction.

Then, our system is used for the QPI of label-free HeLa cell by replacing the objective and the camera with another 0.75 NA objective and 2.2 $\mu$m pixel size camera. The FOV is 285.1 $\times$ 213.8 $\mu$m$^\text{2}$ with the sampling rate of 0.11 $\mu$m in the object plane. Figure 8(a) and (b) show the images of high resolution quantitative phase and the phase gradient in the direction of the image shear (45$^\text{o}$). As can be seen in Fig. 8(c), three sub-regions are selected by solid rectangular shape for no resolution loss of phase images. For this group of quantitative results, We will not repeat the enhancement of resolution of annular LED illumination but point out some defects in quantitative images. The background of this quantitative phase image is not ``black'' enough, which is caused by the loss of low frequency features of Fourier spectrum in the Fourier space. The root cause of this problem is the finite spacing between two adjacent LED elements leading to the mismatching between objective pupil and annular LED pattern. Furthermore, the PTF of system tends to be zero near zero frequency and makes the recovery of low frequency  information difficult.

\begin{figure}[!t]
    \centering
    \includegraphics[width=13cm]{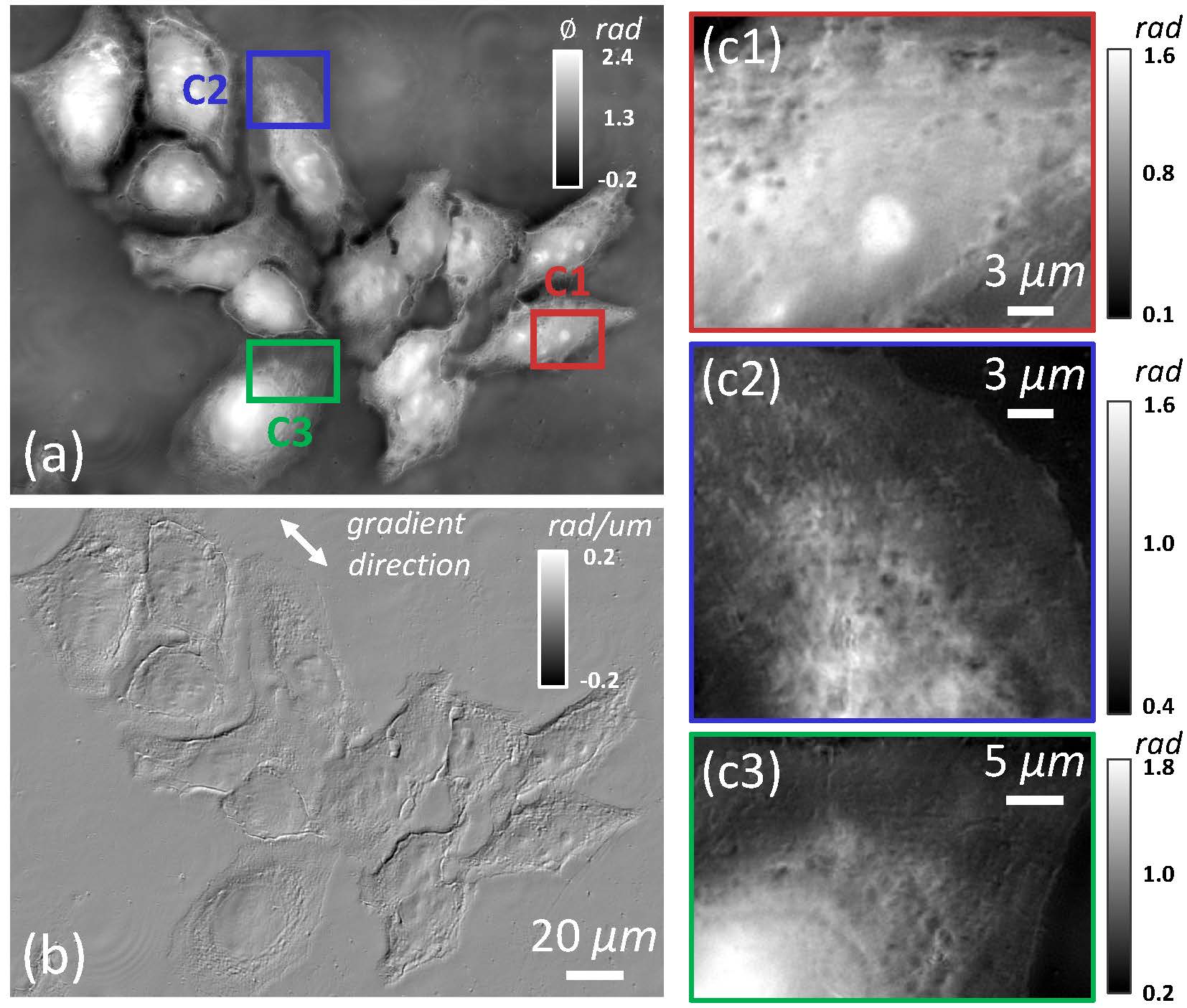}
    \caption{(a) High resolution QPI of HeLa cell with 0.75 NA objective. (b) Simulated DIC image. (c) Three enlarged sub-regions of quantitative phase of HeLa cell. Scale bar equals 20$\mu$m, 3$\mu$m and 5$\mu$m, respectively.}
    \label{}
\end{figure}


\section{Discussion and conclusion}
In summary, we demonstrate an effective QPI approach based on programmable annular LED illumination for twice imaging resolution of objective NA and noise-robust reconstruction of quantitative phase. The WOTF of axisymmetric oblique source is derived using the concept of TCC, and the WOTF of discrete annular aperture is validated with the incoherent superposition of the individual point source. The inversion of WOTF is applied to the intensity stack containing three intensity images with equal and opposite defoci, and the quantitative phase could be retrieved. The recovered phase of simulated resolution target and noise-corrupted test image prove that the proposed illumination pattern could extend imaging resolution to 2 NA of objective and give great noise insensitivity. Further more, the biological sample of human cancer cells are imaged with two different types objective and the imaging resolution of retrieved phase is enhanced compared with the coherent illumination indeed. Besides, this QPI setup is easily fitted into a conventional optical microscope after small modifications and the programable source makes the modulation of annular pattern more flexible and compatible without customized-build annuli matched objective pupil.

But there are still some important issues that require further investigation or improvement in this work. Due to the dispersion of LEDs and the finite spacing between adjacent LED elements, the annular illumination pattern and pupil of objective are not well tangent internally with each other. The unmatched annular aperture with objective may cause the loss of low frequency owing to the overlap and offset of PTF near zero frequency. In other words, the missing of low frequency would lead to that the background of phase images is not ``black'' enough. Another shortcoming of this modified microscopic imaging system is that it is difficult to apply the long-term time-elapse living cellular imaging to these relatively low end bright-field microscope, like Olympus CX22 microscope, different from our early work based on IX83 microscope. To solve these problems, a special sample cuvette is required for the imaging of living biological cells and the additional devices may be needed to modify our setup, such as a smaller spacing and brighter LED array. Despite these existing drawbacks, the configuration of this system takes full advantage of the compatibility and flexibility of the programmable LED illumination and bright-field microcopy. And the annular illumination pattern gives the quantitative demonstration of control samples and promising results of biological specimens.

\section*{APPENDIX}
\subsection*{A. Derivation of Intensity formation under partially coherent illumination using Hopkins' formulae}

In the main text, the standard optical microscope system can be simplified as an extended light source, a condenser lens, a sample, an objective lens, and a camera on the image plane. Based on Abbe's theory \cite{Abbe}, the captured image of object at the image plane can be interpreted as the summation of all the source points of the illumination. For each source point, the optical formation is described by Fourier-transforms and a linear filtering operation as a linear system, and the electrical field $E\left( {x,y} \right)$ on the camera plane can be expressed as
\begin{equation}
E\left( {x,y;{f_c},{g_c}} \right) = \iint { {t\left( {f,g} \right)h\left( {f + {f_c},g + {g_c}} \right)\exp \left[ { - i2\pi \left( {fx + gy} \right)} \right]dfdg}}
\end{equation}
Where $t$ is the complex transmittance of object, and $h$ represents the amplitude point spread function (PSF) of the imaging system. The intensity on the image plane is proportional to the square magnitude of the electric field distribution and takes the form of
\begin{equation}\label{23}
\begin{aligned}
I\left( {x,y} \right) & = {S\left( {{f_c},{g_c}} \right){{\left| {E\left( {x,y;{f_c},{g_c}} \right)} \right|}^2}d{f_c}d{g_c}}\\
 & = {S\left( {{f_c},{g_c}} \right){{\left| {{\mathscr{F}} \left[ {t\left( {f,g} \right)h\left( {f + {f_c},g + {g_c}} \right)} \right]} \right|}^2}d{f_c}d{g_c}}
 \end{aligned}
\end{equation}
Where $I(x,y)$ is the intensity of the object captured at the image plane, $S( {{f_c},{g_c}})$ is the distribution of  extended light source, and ${\mathscr{F}}$ denotes Fourier transform. By interchanging the order of integration, we can express Eq. (\ref{23}) according to Hopkins' formulation \cite{Hopk,MBorn}
\begin{equation}
\begin{aligned}
I\left( {x,y} \right) = \iiiint{} & S\left( {{f_c},{g_c}} \right)P\left( {{f^{'}} + {f_c},{g^{'}} + {g_c}} \right){P^*}\left( {{f^{''}} + {f_c},{g^{''}} + {g_c}} \right)T\left( {{f^{'}},{g^{'}}} \right){T^*}\left( {{f^{''}},{g^{''}}} \right) \\
 & \exp \left[ { - i2\pi \left( {{f^{'}} - {f^{''}}} \right)x - i2\pi \left( {{g^{'}} - {g^{''}}} \right)y} \right]d{f^{'}}d{g^{'}}d{f^{''}}d{g^{''}}
\end{aligned}
\end{equation}
Where $P$ is the coherent transfer function with the objective pupil function $\left| P \right|$, and $T$ is the spatial object spectrum respective to the Fourier transform of object complex transmittance $t$. Here, we separate the contribution of the specimen and system, and the transmission cross coefficient (TCC) is introduced as a combination of the source and pupil expressed as
\begin{equation}\label{25}
TCC\left( {{f^{'}},{g^{'}};{f^{''}},{g^{''}}} \right) = \iint{{S\left( {{f_c},{g_c}} \right)P\left( {{f^{'}} + {f_c},{g^{'}} + {g_c}} \right){P^*}\left( {{f^{''}} + {f_c},{g^{''}} + {g_c}} \right)d{f_c}d{g_c}}}
\end{equation}

By replacing the variable $\left( {{f^{'}},{g^{'}}} \right)$ and $\left( {{f^{''}},{g^{''}}} \right)$ with two 2D vector ${{\bf{u}}_1}$ and ${{\bf{u}}_2}$ in frequency domain, and  the Eq. (\ref{25}) can be simplified as
\begin{equation}
TCC\left( {{{\bf{u}}_1};{{\bf{u}}_2}} \right) = \iint{{S\left( {\bf{u}} \right)P\left( {{\bf{u}} + {{\bf{u}}_1}} \right){P^*}\left( {{\bf{u}} + {{\bf{u}}_2}} \right)d{\bf{u}}}}
\end{equation}
Then, the final intensity of object on the image plane can be rewritten in 2D vector variable
\begin{equation}\label{27}
I\left( {\bf{r}} \right) = \iint{{TCC\left( {{{\bf{u}}_1};{{\bf{u}}_2}} \right)T\left( {{{\bf{u}}_1}} \right){T^*}\left( {{{\bf{u}}_2}} \right)\exp \left[ {i2\pi {\bf{r}}\left( {{{\bf{u}}_1} - {{\bf{u}}_2}} \right)} \right]d{{\bf{u}}_1}d{{\bf{u}}_2}}}
\end{equation}

\section*{ACKNOWLEDGMENTS}
This work was supported by the National Natural Science Fund of China (11574152,
61505081), `Six Talent Peaks' project (2015-DZXX-009, Jiangsu Province, China) and `333 Engineering' research project (BRA2015294, Jiangsu Province, China), Fundamental Research Funds for the Central Universities (30915011318), and Open Research Fund of Jiangsu Key Laboratory of Spectral Imaging \& Intelligent Sense (3092014012200417). C. Zuo thanks the support of the `Zijin Star' program of Nanjing University of Science and Technology.

\section*{DISCLOSURES}
The authors declare that there are no conflicts of interest related to this article.

\end{document}